\documentstyle[12pt,epsf]{article}

\begin {document}
\begin{flushright}
{\small
SLAC--PUB--8468\\
June 2000\\}
\end{flushright}

\vfill\vfill
\begin{center}
{{\bf\LARGE   
The Future of Particle Physics}\footnote{Work supported by
Department of Energy contract  DE--AC03--76SF00515.}}

\bigskip
James D. Bjorken\\
{\sl Stanford Linear Accelerator Center \\
Stanford University, Stanford, California 94309 USA\\
E-mail:  bjorken@slac.stanford.edu}
\end{center}

\vfill

\begin{center}
{\bf\large   
Abstract }
\end{center}

After a very brief review of twentieth century 
elementary particle physics, prospects for the next century are discussed. 
First and most important are technological limits of opportunities; next, the 
future experimental program, and finally the status of the theory, in 
particular its limitations as well as its opportunities.

\vfill

\begin{center} 
 Invited talk given at the  \\
     International Conference on Fundamental Sciences: \\
     Mathematics and Theoretical Physics \\
Singapore\\
13--17 March 2000\\
\end{center}

\vfill
\eject

\baselineskip=14pt

\section{Preamble}

It is a real pleasure to be back in Singapore and have the opportunity to
address this meeting. About a decade ago, I gave a similar talk here, as a
last-minute substitute for Leon Lederman~\cite{ref:a}. Since that time I have for the
most part exited theoretical physics and entered experimental physics. In
particular I have been promoting new detector techniques for exploring the
world of strong interactions. This led to my colleague Cyrus Taylor, a
string theorist by training, and I leading a small test/experiment in the
Fermilab Tevatron Collider. The results of this experiment were
modest~\cite{ref:b}. But the experience was in many ways extremely rewarding and
enriching, and in other ways frustrating. However, as the new century
begins, I have emerged from the experimental trenches and am entering the
world of retirement. I now do have the opportunity to again take a look at
the big picture, after a decade of time off. And the invitation to give this
talk provided a most opportune way to organize my own thoughts and to try to
express them.

\section{The Big Picture}

Twentieth century physics featured several great syntheses, The first was
the extraordinary synthesis by Planck of thermodynamics with Maxwell
electrodynamics, giving birth to quantum theory. It was quickly followed by
the synthesis of classical mechanics with electrodynamics by Einstein,
giving birth to special relativity.  And in the 1920s, with the decisive
synthesis of Newtonian mechanics with the ``old" quantum mechanics by
Heisenberg and Schrodinger, there also emerged quantum electrodynamics (or
QED), the synthesis of Maxwell electrodynamics with quantum mechanics.

All this happened in less than three decades. Much of the remaining history
for the century belongs to the experiments, which built upon these
foundations and in particular drove the development of particle physics.
Many particles were discovered.  The strong and weak forces themselves
needed to be discovered before there was any opportunity to understand their
role. Eventually things became understood well enough that by the 1970s
there was the possibility of a synthesis of weak and electromagnetic forces
into a common structure: the SU(2) $\times$ U(1) electroweak component of the
Standard Model. And the strong force is now successfully described by the
generalization of QED called quantum chromodynamics (QCD), based on the
exact SU(3) color symmetry of quarks and their force carriers, the gluons.

All these forces (including gravitation) are described at short distances
by the same class of fundamental theories, the so-called gauge theories.
The force law at short distances is in all cases essentially
inverse-square. Standard Model forces are proportional to a variety of
conserved charges, while gravity couples to energy-momentum. So as we
enter the twenty-first century, there is a strong anticipation that these
four forces have a common origin and that further synthesis is on the way.

\begin{figure}[htb]
\begin{center}
\leavevmode
 \epsfbox{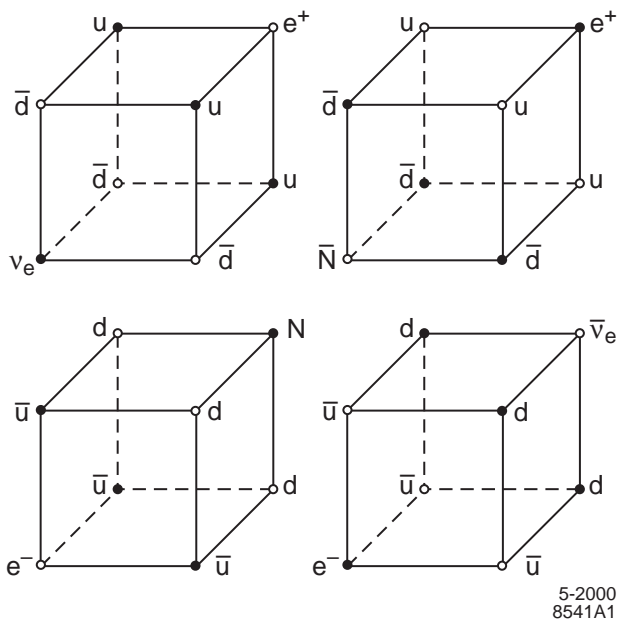}
\end{center}
\caption[*]{Building-blocks of the Standard Model.  (The solid dots comprise
the SO(10) {\bf 16} for the two-component, left-handed fermion degrees of
freedom; the open dots comprise the $\overline{\bf 16}$ antiparticle representation.)}
\label{fig:1}
\end{figure}

There are quite specific clues to build upon. One clue is the pattern of
fundamental building blocks of matter: the quarks and leptons.  They are
exhibited explicitly as building blocks in Fig.~\ref{fig:1}. This is a construction
which has technical meaning only when viewed beyond the Standard Model, when
the SU(3) $\times$ SU(2) $\times$ U(1) Standard Model symmetry group is embedded in the
much larger symmetry group of ten-dimensional rotations, SO(10). A clumsy,
reducible representation of the Standard Model group becomes an elegant,
single sixteen-dimensional fundamental spinor representation of SO(10). In
that context the 5-dimensional cube which is depicted in Fig.~\ref{fig:1} has a
definite mathematical meaning.

\vspace{.5cm}

\begin{figure}[htb]
\begin{center}
\leavevmode
\epsfbox{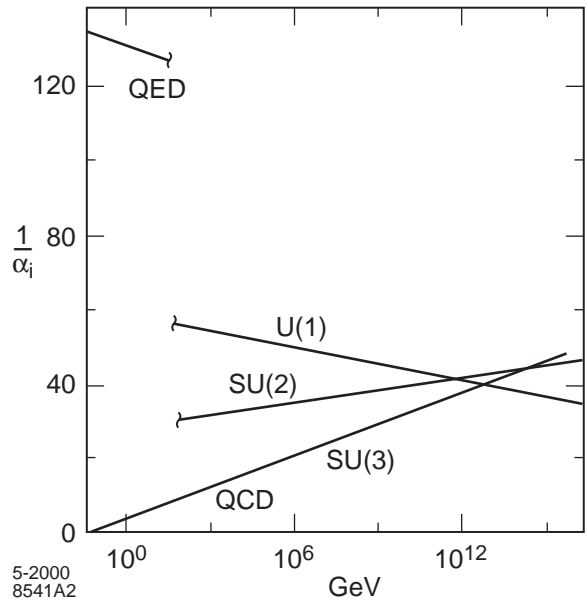}
\end{center}
\caption[*]{The ``fine structure constants'' of the basic forces as
function of momentum scale.  Note the discontinuity in the electromagetic
coupling strength at the electroweak scale, due to the replacement of
$U(1)_{CM}$ with the Standard Model $U(1)_Y$.}
\label{fig:2}
\end{figure}

A second clue, pointing in the same direction, has to do with the fact
that the coupling strengths or ``charges" associated with the three kinds
of Standard Model ``gauge" forces vary slowly with distance scale. It
happens that they converge (or very nearly converge) to a common value at
a very high mass scale, 10$^{15}$ GeV, give or take a factor 10, as shown in
Fig.~\ref{fig:2}. It is at this scale that synthesis of these three forces can be
expected to occur. This anticipated synthesis is known as Grand
Unification.

Even one or two generations ago, it was the dream of every theorist to
come to an understanding of the value of the pure number 1/137 of QED, a
number which characterizes the intrinsic strength of the electromagnetic
force at the quantum level. This is no longer the case. The 137 evolves to
128 through ``vacuum polarization" effects; cf. Fig.~\ref{fig:2}. At this point, at a
mass scale of about 100 GeV, the electroweak synthesis takes place.
Thereafter it is the coupling strength of a mixture of photon and
electroweak boson which, together with weak-boson and gluon coupling
strengths, evolve from values of about 1/60, 1/30, and 1/10 respectively
to the common value of about 1/40 at the grand-unification scale. So the
137 has been divided by a factor of about 3 1/2, and the question is now
``Why 1/40?"

\vspace{.5cm}

There is another synthesis taking place at present, that of cosmology and
particle physics. Twentieth century astronomy and astrophysics has its own
rich history. But with the emergence of Big Bang cosmology in the last
forty years, there has also emerged an increasingly strong interdependence
of cosmology and particle physics. The high temperatures present during
the early epochs of the Big Bang demand a good understanding of particle
physics in any theoretical description, while the empirical evidence which
constrains theories of the early history of the universe also constrains
the theories of particle physics.

\begin{figure}[htb]
\begin{center}
\leavevmode
\epsfbox{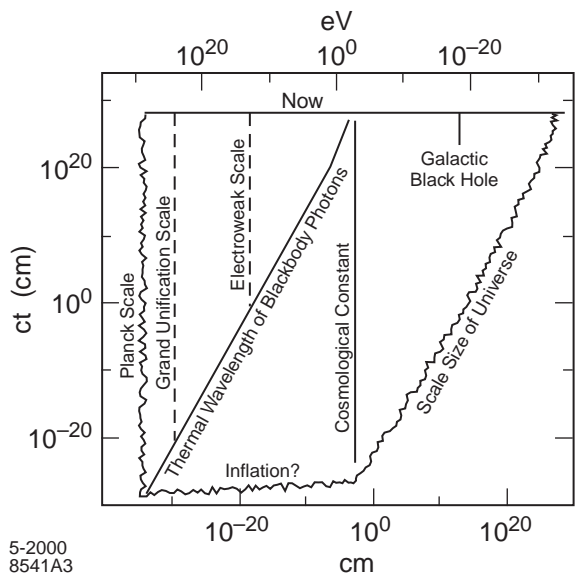}
\end{center}
\caption[*]{The Big Picture:  important values of distance (or momentum)
scales versus time (as measured from the Big Bang).}
\label{fig:3}
\end{figure}

So there is now an increasingly strong interconnection of the largest
distances with the smallest, and of the longest time scales with the
shortest. This is illustrated in Fig.~\ref{fig:3}, which can truly be called the Big
Picture. In that figure every effort has been made to omit the
superfluous. What is left is what I personally believe to be the key
landmarks necessary to comprehend when moving on to the next syntheses. 

\vspace{.5cm}

Nothing of the Big Picture in Fig.~\ref{fig:3} existed a century ago. That it
exists now is a tribute to the extraordinary scientific progress made in
that period. Progress in physical science has three components:
technological, experimental, and theoretical. These are interconnected,
but I believe that the order of importance is as stated. Without
technological advances, experimental technique stagnates. And without the
validations and unanticipated discoveries that comes from advances in
experiment, the finest creations of theoretical physics languish as
exercises in natural philosophy or in higher mathematics, and are of
little worth as physical theory.

\section{Technology}

In particle physics the name of the game is energy. Throughout the
twentieth century there has been exponential growth in the attainable
center-of-mass energy available for particle collisions, beginning with a
few electron-volts in early vacuum tubes and ending with the trillions of
electron volts in the Fermilab Tevatron Collider~\cite{ref:c}. On average this amounts
to a doubling time of about 2-1/2 years.

This pace is unlikely to be equaled in the coming century. The slowing of the
pace has already been apparent for the last decade or two. The new machines
are big and expensive, and take a long time to build and to fully exploit. But
this does not mean an end to the field.  There is plenty of room for
further expansion of the possibilities, throughout the coming century.

Modern colliding beam machines are circular storage rings, within which
some combination of counter-rotating beams of electrons, positrons,
protons, and/or antiprotons collide with each other.  The biggest
electron-positron collider is the LEP ring at the CERN laboratory~\cite{ref:d} in
Geneva, Switzerland, 27 kilometers in circumference, and now running at a
cms energy of about 200 GeV. This represents the limit of possibility for
this kind of ring. The electrons and positrons emit so much synchrotron
radiation, and the rate grows so rapidly with increasing energy, that it
is impractical to contemplate similar machines at much higher energies.  
This is not the case for the heaver protons, where ten times that energy
is attained in the Fermilab proton-antiproton collider already. And the
Large Hadron Collider (LHC), under construction at CERN in the LEP tunnel
itself~\cite{ref:e}, will increase this value sevenfold. Sooner or later, a similar
synchrotron-radiation limit will be reached, however, especially given
that superconducting magnets are needed to bend the protons, and that
superconducting systems and intense beams of x-rays tend not to peacefully
coexist. A rough estimate of where the ultimate limit for circular proton
rings occurs puts the number at a few hundred TeV per beam. This is between
one and two orders of magnitude larger than the LHC, and implies that at
least one more very big collider is technically feasible~\cite{ref:f}, and perhaps
affordable on a world scale sometime in the future.

Such a machine would be very big, probably over a thousand kilometers in
circumference. It would require a lot of (underground) real estate, and
for a long time it has seemed to me that this region of the planet, with
Down Under so close at hand, is a natural focus when dreaming about this
possibility.  Given the premise that the twenty-first century will witness
not only an economic but also scientific and cultural blossoming within
Southeast Asia, such a project in this region might be an appropriate and
locally beneficial way of entering this fundamental field of science. The
time scale for even considering such an initiative is not at all
immediate, but could be as short as two or three decades. And as we shall
see, there are scientific considerations as well to deal with. What is
done in the far future depends in an important way on what will be learned
from the experiments performed in the nearer future.

When thinking about higher energy electron-positron collisions, the
technique of choice is to accelerate the beams within straight-line,
linear accelerators, as done in my home institution SLAC at Stanford, where
collisions with 50 GeV electrons and positrons per beam have been
attained~\cite{ref:g}. Another factor of 10 to 20 can be contemplated using relatively
straightforward extensions of existing techniques.  The main issue is to
do this in an energy-efficient, economical way. A great deal of study for
the ``Next Linear Collider" (NLC) is being carried out, with a large amount
of international collaboration. And some kind of NLC is a leading
candidate, perhaps the leading candidate, for the next large facility
beyond the LHC needed to further push back the high energy frontier~\cite{ref:h}.

\begin{figure}[htb]
\begin{center}
\leavevmode
\epsfbox{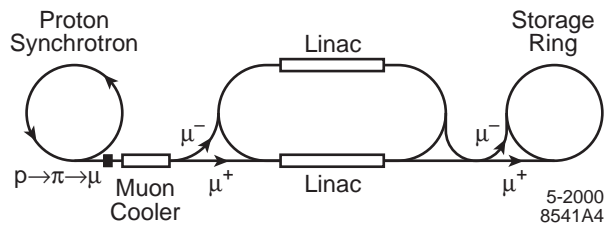}
\end{center}
\caption[*]{A schematic of a muon collider.  A very intense proton source
creates a beam which is extracted and targeted.  The pions and decay muons are
focussed and collected.  The large phase space occupied by the muon beam is
diminished by ``ionization cooling".  Then the muons are accelerated in a racetrack
linac before being injected into the storage-ring collider.}
\label{fig:4}
\end{figure}

However there is a relatively new, competitive idea being studied as well,
namely to collide muons and antimuons (the unstable, heavier
second-generation copies of the electrons and positrons) with each other
within circular storage rings (cf. Fig.~\ref{fig:4}).  At first sight this idea
looks crazy, but after careful scrutiny it becomes less crazy, albeit
difficult and adventurous~\cite{ref:i}.  
One physics advantage of muons over electrons and positrons is that resonant
production of a Higgs boson is much larger and possibly observable.
Another advantage is that muons do not emit so much synchrotron radiation. However
many of them do decay en route to the collision point, and their decay
products create a serious nuisance.  An extremely intense source of primary muons
is required.

\vspace{.5cm}

But a general advantage of a muon collider facility is that there are many
side-benefits. One can easily see that each component of the facility
can---and should---support other physics programs. A very intense source of
protons of tens of GeV is required, and this by itself is a good device
for studying properties of the hadrons, including {\it e.g.} the rare decays of
kaons. And the muons are accelerated in a recirculating linear accelerator
which is a more powerful and sophisticated version of a facility, CEBAF,
now being used to study hadron structure via electron scattering~\cite{ref:j}. Perhaps
the most promising secondary application is utilization of the intense
beams of neutrinos created by the decaying muons. Just in the last few
years, with the emergence of the evidence for neutrino mass and mixing
from non-accelerator experiments, there has been an escalating interest
within the particle accelerator community in creating a ``neutrino factory"
using muon-collider technology.  Such a program would provide useful
neutrino physics as well as providing proof-of-principle evidence that the
idea of creating intense, bright muon beams really can be made to work~\cite{ref:i}.

So while consideration of an NLC, with center-of mass energy of up to 1--2 TeV,
will be high on the options list of future facilities beyond the LHC, 
the muon colliders, which might reach to a few TeV in the center
of mass were all to go well, will also be under consideration. It does have to be said that the
muon-collider technology is new and untested, so that it is reasonable to
expect that if that route is followed it will be considerably further into
the century before one could contemplate high energy muon collisions being
attained. In either case, the reason that such lepton colliders are
competitive with the much
higher energy $pp$ colliders, such as the LHC, is the efficient conversion
of all the energy in each event into interesting physics.  The advantage
of higher energy in $pp$ colliders is mitigated by the complexity of the
collisions and the relative rarity of the collisions which produce the
highest-priority ``new physics".

In addition to the energy frontier, there will be many other frontiers
remaining within particle physics. There are good reasons for producing
interesting particles with the highest intensities possible. Accelerator
facilities which specialize in production of a single kind of particle are
already in abundance: we already see $B$ factories, $Z$ factories, $K$
factories, and $\phi$ factories. And as mentioned above, we probably will see
neutrino factories in the future. Once the Higgs sector, the prime target
of experimental physics nowadays, is found, there will be an irresistible
urge to create a Higgs factory as well, probably utilizing some kind of
lepton-lepton collider. It is likely that the ``factory" programs will have
great longevity.

Another frontier is that of non-accelerator facilities, utilizing for
example nature's own particle beams, be they cosmic rays or neutrinos from
the sun. When the ideas of grand unification emerged (Fig.~\ref{fig:2}), there also
emerged the prediction of proton decay. This was a real turning point for
the field of non-accelerator particle physics, because thereafter the large-scale
investment in huge detectors became commonplace, and the level of
sophistication relative to previous, traditional cosmic ray research
escalated considerably. Even today the investment in non-accelerator
particle physics is a relatively small fraction of the total, so that
important measurements may be expected in the future to be approached on a
scale larger than done at present. The recent advances in neutrino
physics, especially in Japan with SuperKamiokande~\cite{ref:l}, bear witness to the
value of making a large, sound investment in a good detector when the
potential benefits warrant it.

But the non-accelerator experiments will forever be very difficult and
fraught with more uncertainty that the more highly controllable
accelerator experiments. So there will be strong motivation to push the
high energy frontier well beyond what we have talked about so far. It is
here that very difficult barriers appear.

If one wants to attain center-of-mass energies well beyond 1000 TeV, there
seems to be very little choice but to do it with linear acceleration. If  we ask 
to do this within  a reasonable distance, say 100 km, this
implies an average acceleration gradient of at least 10 GeV per meter, or
1 eV per Angstrom. High powered lasers can in principle provide forces of
such a magnitude. But such forces, if present in matter, are strong enough
to pull electrons out of metals, or in general destroy the orderly
chemical structure of just about any material.  There is a general
rule, Lawson's theorem, that states that free electromagnetic waves do not
make a good accelerator, essentially because the electric vector is at
right angles to the Poynting vector.

So a good accelerator will almost certainly have a material structure
close to the beam. or perhaps a plasma within the beam. And this material
structure might well be damaged or destroyed by the laser pulse or other
source of the accelerating field every time the device is pulsed. I
envisage this as the insertion of an accelerating structure into the
beam, instead of the present practice of insertion of a beam into the
accelerating structure.

The notion of an accelerator which is destroyed by the beam during every pulse is
not necessarily a hopeless one, especially if the transverse dimensions of
the machine are kept very small, not much larger than the beam itself. The
small transverse dimensions are probably a necessity anyway on grounds of
the energy budget. One cannot afford to fill a large cavity with a lot of
energy, as presently done, especially when it does not directly contribute
to the acceleration of the beam.  So it could be that the accelerating
structure is in the form of a sequence of structured tapes, or edges of
structured rotating disks, which move through the beam region. The
transverse scale might be anything from microns to millimeters.

And there is a new technology, nanotechnology, which although not viable
now, might in the future evolve to the point that accurate, practical, and
inexpensive ({\it i.e.} disposable!) acceleration microstructures might
eventually be fabricated. Nevertheless, the trouble list associated with
such a line of thinking is impressively formidable. A huge obstacle is
that of maintaining the brightness of the beam, {\it i.e.} keeping the beam size
small. The typical unit of periodicity (cell length) of such a
``miniaturized" linear accelerator might be envisaged to be in the range of
millimeters at most. So in a 100 km machine, there are at least 10$^8$ such
cells. And the phase-space volume of the beam (emittance) cannot increase
on average by much  more than 1 part in 10$^8$ per cell or it will grow unacceptably
large by the end of the machine.  So the reproducibility of these
extremely violent acceleration mechanisms must be maintained to a very
high level of accuracy. I am sure this is only one of a number of
similar problems.

There is in addition another great difficulty, which is motivation. It is
a folk theorem in the trade that a project does best if there is strong
physics motivation for it. The sense of urgency creates focus, drives the
project forward, and encourages everyone to contribute that extra level of
creative effort to make it successful. An R\&D program on very high
gradient linear accelerators will naturally be restricted to short prototypes
for a long time, and therefore not be highly physics driven.
And the primary physics demand is not only for high energy
but also for an extremely high collision rate. Typically, for each factor
of ten of energy increase, one needs to increase the luminosity, or
collision rate, by a factor of order one hundred. But there are secondary
physics topics which can be addressed. For example, were 10 GeV electron
or proton beams to be produced, economically, on a tabletop in the
basement of some university physics department, some physics uses for them
might well be found. This is especially true for proton beams, where many
physics applications do not require the relatively large intensities that
electron-beam physics requires.

\eject

At present, there is actually considerable activity in the field of high
gradient linear acceleration, not only theoretical but increasingly
experimental~\cite{ref:m}. It seems to me that there should be a
high level of attention paid to this problem, even though it does not
look to be immediately practical, and even though super-high gradients do not
seem to be needed for at least a couple of generations of machines, in
order to push the energy frontier forward. Sometime in the future the
field will have to face up to this ultimate problem, and doing the
homework now would seem to be a prudent strategy.

\section{Experiments}

In the realm of accelerator-based particle physics, the technology of
particle detectors is highly advanced. In the future, as prognosticated  in the
previous section, refinements and extensions of the existing detector
technologies can be expected to occur, especially with regard to rate
capability, pattern recognition, resolution, and management of increasing
data volume. But I at least do not perceive any need for a fundamental
change of approach. The most difficult frontier is the providing of the
higher energy collisions themselves, not in detection and analyzing the
collision products.

Perhaps the most interesting experimental frontier is sociological. The
size and complexity of experimental collaborations continues to grow. At
present we see, at the LHC, collaboration sizes approaching two thousand,
with participation of hundreds of collaborating institutions around the
globe. And the time scale for construction and full exploitation of a
single detector for a single experimental program rivals the full career
length of the participating physicists. When I was here in Singapore the
last time, I devoted quite a lot of time to this issue~\cite{ref:a}. The social
problems are still there. But there remains a large, enthusiastic young
generation within the big detector collaborations, and I believe that the
system, although very different from what it was two or three decades ago,
still allows a highly creative and stimulating scientific environment.

Another singular feature of modern experimental particle physics is its
professionalism. In the past two decades, the standards applied to all
aspects of the experiments, especially with regard to the management and
statistical analysis of the data, have soared. This is I think largely due
to the size of the collaborations: they are large enough to contain
experts in all the relevant subspecialties. I think these standards are
well above what exist in other fields of science, and that participants
in big experiments, even if they leave the field of particle physics at
some point, are superbly prepared to apply that expertise in other
scientific or technological applications.

Indeed, it must not be forgotten that experimental particle physics
not only depends upon technological progress, but that it contributes to it.
For example,  it was the need for efficient communication  and for transmission 
of very large volumes of information amongst the highly
dispersed international community of collaboration members that gave birth
to the World Wide Web at CERN~\cite{ref:n}. It is a perfect example of how research in basic
science is not only an intellectual endeavor worthy of support by society
for its own sake, but also an engine that drives economic progress in ways
that create fundamental changes which are not programmable in advance.

It therefore seems to me that it is very important for Singapore and this
region to not only consider the enhancement of its participation in basic
science in the areas of theory, as represented in this meeting, but also to
consider participation in  experimental and accelerator
physics. It is especially easy nowadays to enter the field via the large
international experimental collaborations and thereby gain immediate
access to the mainstream of the field. The benefits grow in proportion to
the investment, and I would urge the creation of conditions that might
allow participation of especially the younger generation in
particle-physics experimentation.

\section{Beyond the New Standard Model}

We now turn to the theoretical situation. I believe, as do many others,
that the most crucial problem we face has to do with understanding the mechanism
by which the quarks, leptons, and electroweak force-carriers (the gauge bosons
$W$ and $Z$) get their mass. This problem is characterized by a mass scale at
the edge of measurability at present, of order 100 to 1000 GeV. It was
well defined already twenty years ago, and the efforts of many
experiments, which have included the discovery and measurement of the
properties of the $W$, $Z$, and the sixth, remarkably massive top quark,
have simply sharpened the basic issues.

But the situation regarding the Standard Model has not remained static, thanks
to the newest generation of neutrino experiments.  They provide strong
evidence that neutrinos also have mass and undergo ``mixing", similar to the way
the three generations of quarks are mixed by the electroweak interactions.  This
is an extremely significant discovery, although not really a revolutionary one.
It is in fact a very welcome one, one which replace the Old Standard Model
by a New Standard Model.  The 20 or so free parameters of the Old
Standard Model are now increased by at least 7 (3 neutrino masses and
4 mixing angles) and perhaps 12 new parameters (masses for the 3 $N$'s 
discussed below plus two other mixing angles), depending upon what one
chooses to include in the count. This by itself may not seem like progress.
 But the reason that neutrino mass is not all
that unwelcome has to do with the Grand Unification perspective we mentioned
earlier.  In the Old Standard Model, there are 15 spin-1/2 building blocks per
generation:  the sixteenth degree of freedom $N$ in Fig.~\ref{fig:1} is not included.  The
natural GUT symmetry group to consider is SU(5), for which the 15 building
blocks fill two multiplets: a \boldmath $5$\unboldmath\ and a \boldmath${\overline{10}}$.
\unboldmath
Neutrino mass, in particular the very small  neutrino mass observed,
occurs in the SO(10) extension when the extra degrees of freedom $N_i$
(one for each generation of fermions) possess very large masses of order of the
10$^{15}$ GeV GUT scale.  The ordinary neutrinos mix with the $N$'s, and
quantum mechanical level repulsion drives their masses to very small values
in a natural way.  And with the $N$'s present, the pattern of the building blocks
(Fig.~\ref{fig:1}) becomes simpler, as mentioned before, being described by a single
16-dimensional irreducible spinor representation of SO(10).  So the New
Standard Model is to the Old Standard Model as SO(10) is to SU(5):  a little
bigger, and also a little prettier.  And the transition from ``Old" to ``New"
should be regarded as a very strong clue in dealing with the question of how
to synthesize the strong and electroweak forces at the GUT scale.

Nevertheless, the problem of how the masses of quarks, leptons, $W$,  $Z$, and neutrinos,
are generated still must be faced.
There are many ideas on how the problem of mass is to be solved. The most
economical way requires only one extra particle to exist, the famous Higgs
boson.   In many ways,
this simplest of scenarios, called the ``desert" scenario, works the best.   Consistency
is maintained all the way to the GUT scale
of 10$^{15}$ GeV or so if and only if the mass of the Higgs boson lies in a narrow
window: 160 $\pm$ 20 GeV.  Nevertheless, theorists remain very disquieted by
the ``desert" scenario because there appears a quadratic divergence in the Higgs
mass which must be subtracted away, and it seems very unnatural to set the
physical Higgs mass to a small value at the electroweak scale of 100 GeV or 
so, rather than at the GUT scale of 10$^{15}$ GeV.  

However, this
quadratic divergence problem bears great similarity to the problem of
the small value of the cosmological constant.  It too suffers from power-law
divergences.  And the cosmological constant problem cannot be solved with a
straightforward appeal to supersymmetry, as is commonplace for the Higgs
problem.  So a serious ``solution'' to this ``hierarchy'' problem is to ignore it,
acknowledging that its resolution will require a much deeper level of 
understanding, but recognizing that in the meantime renormalization (subtracting
out the infinity) suffices to remove the problem from all known phenomenology.

But angst over the ``hierarchy problem" has in the past 25 years spawned alternative
approaches, the most important of which are phenomenological supersymmetry and
``technicolor".  Both are characterized by the introduction of many new particles
as well as new forces~\cite{ref:o}.  If either is correct, there will be an extraordinarily rich
experimental program for the future.

The evidence from precision measurements of electroweak processes, especially at
LEP, seems to favor a Higgs sector of relatively low mass, below 250 
GeV~\cite{ref:o}.  This
in turn is more in line with the supersymmetry option (``Minimal Supersymmetric
Standard Model", or MSSM), or the ``desert'' scenario than with technicolor,
although nothing is strictly ruled out.  The MSSM is expecially rich in new particles
and new parameters.  Each known particle has its superpartner, differing in spin
by 1/2, most of which having masses below a few TeV.  There are about
100 extra parameters to be determined through experiment.  The Higgs sector is 
bigger, and it is expected that at least one of the Higgs' lies below 130 GeV.

From the experimental point of view, these two alternatives, MSSM and ``desert", stand in 
stark contrast to one another.  If MSSM is correct, then---once the energy scale has
increased sufficiently to produce this cornucopia of new particles and phenomena---it will
be full employment for experimentalists (as it already has been for the MSSM theorists).

\begin{figure}[htb]
\begin{center}
\leavevmode
\epsfbox{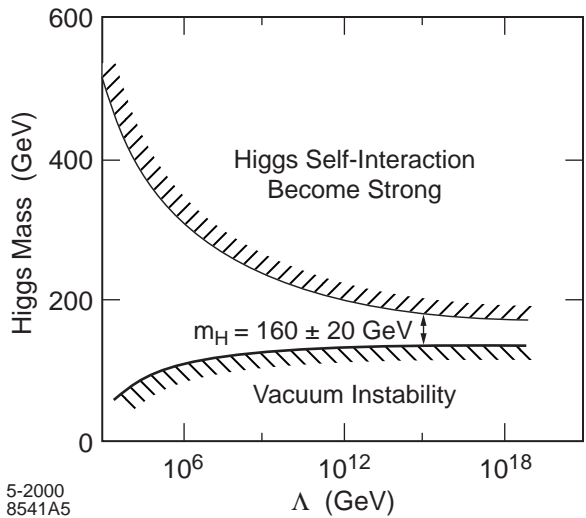}
\end{center}
\caption[*]{Values of $m_{\rm Higgs}$ and momentum scale for which the Standard
Model exists, {\it i.e.} where electroweak perturbation theory converges.  The
upper region is forbidden because the self-interactions of the Higgs particle become
strong.  The lower region is forbidden because the vacuum itself becomes unstable.}
\label{fig:5}
\end{figure}

On the other hand, suppose none of that is found and only the single Higgs boson is seen,
with a mass of 160 GeV as {\it predicted} by the ``desert'' scenario (cf. Fig.~\ref{fig:5}).  It will
be more difficult to motivate new, expensive facilities at higher energies if no clear
landmarks for new phenomena exist.

\vspace{.5cm}

\begin{figure}[htb]
\begin{center}
\leavevmode
\epsfbox{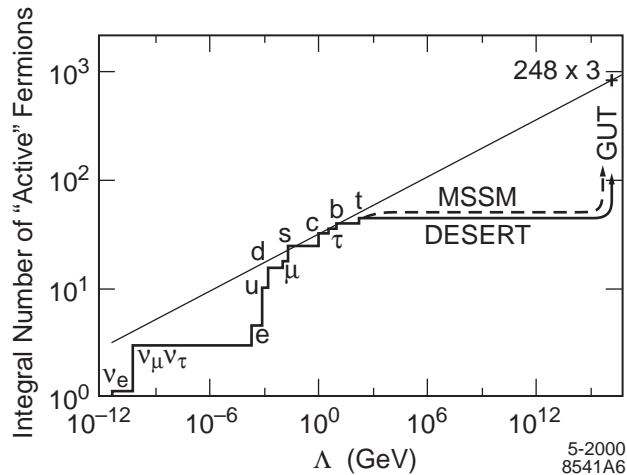}
\end{center}
\caption[*]{Number of (Weyl) spin 1/2 particles (not including antiparticles) with
mass less than $m$, versus $m$, for the ``desert'' scenario and the MSSM.  At
or beyond the GUT scale, many theories anticipate this number to be many
hundreds to above a thousand.}
\label{fig:6}
\end{figure}

But the MSSM and ``desert" scenarios are extreme cases.  Yet another way of dealing with
the hierarchy problem is simply not to have a hierarchy at all, but to have many new
mass scales for new physics between the electroweak and GUT scales.  In Fig.~\ref{fig:6}
is plotted versus mass $m$  the number of (2-component) spin 1/2 fermions possessing
mass no larger then $m$.  At the GUT or Planck scale many theories, {\it e.g.} superstrings, 
end up with many hundreds of fermions.  For example, three E(8) generations  adds up to 744
fermions.  At present energies we have $\sim$ 48.  In both the ``desert'' scenario and the
MSSM nothing much is supposed to happen across all those orders of magnitude.  But
perhaps there are ``oases'' of new physics all across the desert, which gradually release
new degrees of freedom.  The hierarchy problem becomes irrelevant.  But, just like the
``desert'' scenario for experimentalists, the ``oasis'' scenario is a plague for theorists, because
the vision of the GUT scale becomes obscured by everything which is in between, 
most of which is beyond experimental access.

\vspace{.5cm}

The bottom line is simple.  The next generation of experiments is sure to be a singularly
important turning point for the field.  The importance of the Higgs and new particle
searches cannot be overrated.

\section{Boundaries of Knowledge and Theories of \\ Everything}

For the last two decades, the subfield of theoretical particle physics with the most vitality,
and with the most powerful intellectual force applied to it, is undeniably that of string theory.
It is a most ambitious subfield, with the often claimed goal to be no less than a
``theory of everything", one which addresses the ``unsolved'' problem of synthesizing
quantum mechanics with general relativity~\cite{ref:q}.

While I have no problem with what people do, I do have a problem with the rhetoric.  In
my opinion, a ``theory of everything" is not a subfield of physical science, where a theory
requires validation by experiment, but rather is a subfield of ``natural philosophy",
which includes such fields as mathematics, philosophy, and religion, and which allows
speculations and investigations unfettered by the constraints of experimentation
and of the scientific method.

I also question the assertion that we presently have no quantum field theory of gravitation.
It is true that there is no {\it closed}, internally consistent theory of quantum gravity
valid at all distance scales,  But such theories are hard to come by, and in any case, are
not very relevant in practice.  But as an {\it open} theory, quantum gravity is arguably
our {\it best} quantum field theory, not the worst.  Feynman rules for interaction of
spin-two gravitons have been written down, and the tree-diagrams (no closed loops) provide
an accurate description of physical phenomena at all distance scales between
cosmological scales, down to near the Planck scale of 10$^{-33}$ cm.  The
divergent loop diagrams can be renormalized at the expense of an in-principle infinite
number of counterterms appended to the Einstein-Hilbert action.  However their effects
are demonstrably small until one probes phenomena at the Planck scale of distances and
energies~\cite{refnew}.  

One way of characterizing the success of a theory is in terms of
bandwidth, defined as the number of powers of ten over which the theory is
credible to a majority of theorists (not necessarily the same as the
domain over which the theory has been experimentally tested). From this
viewpoint, quantum gravity, when treated---as described above---as an effective
field theory, has the largest bandwidth; it is credible over 60 orders of
magnitude, from the cosmological to the Planck scale of distances. The
runner-up is QCD, which loses credibility at the GUT scale.  Above that
scale QCD arguably gets synthesized into a new improved theory, in a way
perhaps similar to the way QED gets synthesized at the electroweak scale.
Indeed of the three theories, QED as formulated by Dirac and Heisenberg
and renormalized by Feynman, Schwinger, and Tomonaga, has the worst
bandwidth because it is already modified in an essential way at the
electroweak scale.

In the old days QED was considered by Landau and others as an inconsistent
(closed) theory, because the coupling constant $\alpha$ grows at short
distances and eventually, at an incredibly short distance, blows up. This
would in principle limit the bandwidth of QED to a mere 100 orders of
magnitude or so. What a disaster! But in fact other physics intervenes. It
is interesting that this inconsistency does not happen in QCD. The strong
coupling constant only blows up in the infrared, where---with the help of
experimental evidence---it is concluded that the theory remains consistent.
Nevertheless, while QCD does enjoy the (unique) status of an
experimentally relevant and logically consistent quantum field theory with
infinite bandwidth, in practice it probably does not matter that this is
the case, because almost certainly new physics will intervene at the GUT
scale, if not sooner.

While quantum gravity may have splendid bandwidth, it still remains
the case that at the Planck scale the effective field theory formalism
totally falls apart. It is here of course that one finds the arena
appropriate to the Theories of Everything. It is to be sure a valid and
important arena. And it will be wonderful indeed if a successful theory
can be put together at that scale. The trouble with doing so is that there
is precious little guidance from experiment. What has always been typical
for progress in physical science is a painful, slow progression, one step at a time,
with guidance from experiment at most of the steps. Success with a theory
of everything would be something very different and extraordinary---that of
a theory found almost completely using arguments of symmetry
and/or esthetics.

\eject

But no matter what, there are already many beneficial consequences of the
theory-of-everything program. In its present state I see it as a theory of
theories--the study of deep connections between different beautiful
theories, some of which might conceivably be relevant to real physical
phenomena. These connections will I am sure be covered in other talks.
Certainly the theoretical phase-space of ideas has been greatly enlarged,
for example in the considerations of supersymmetries, of extra space-time
dimensions, of black hole phenomena, and even of more efficient ways for
calculating Feynman diagrams. I would be very surprised if the future,
improved theory does not contain some of the ideas spawned by the
superstring revolution, even if superstrings have nothing to do with
anything.

\section{Macroscopic Quantum Gravity}

Given the argument that quantum gravity is a good theory
because it has large bandwidth, I now worry about whether I believe it.
The issue is only whether the quantum-gravity phenomena not covered by
perturbative Feynman-diagram calculations (which I consider safe
territory) rest on solid foundations. This boils down to the question of
black hole horizons and Hawking radiation, a subject which involves a
nontrivial application of quantum gravity at distance scales large
compared to the Planck scale. (The physics near a true gravitational
singularity will also be nontrivial, but will be characterized presumably
by physics at the Planck scale.) The potential problem that concerns me is
that the black hole horizon (Schwarzschild radius) is a region
characterized by very large complexity and very large bandwidth.

By this I mean the following. In Schwarzschild coordinates, place
stationary observers a very small distance $h$ above the horizon. Their job
is to survey the local environment, as well as communicate with neighbors.
As the horizon is approached, the spacing of such observers must decrease
as $\sqrt h$; otherwise they will not be able to send light signals to 
neighboring stations; the photons will fall into the black hole before
getting there. So the number of such stations must be proportional to
the black hole area, and depend inversely upon the height above the
horizon. This result no doubt has something to do with black hole entropy,
and is what I mean by increasing complexity as the horizon is approached.

The increase of bandwidth is related to the fact that, because of
gravitational redshift,  the surveyor's clock
rate decreases toward zero as the horizon is approached,
implying infrared sensitivity. The surveyor will
see a divergent ratio of frequency of the light from his own Cesium clock,
relative to light received from a distant Cesium-clock frequency standard.
In addition the surveyor gets hot; he feels himself immersed in local
Hawking radiation, with Hawking temperature which again diverges as the
horizon is approached.

\begin{figure}[htbp]
\begin{center}
\leavevmode
{\epsfysize=4.45in\epsfbox{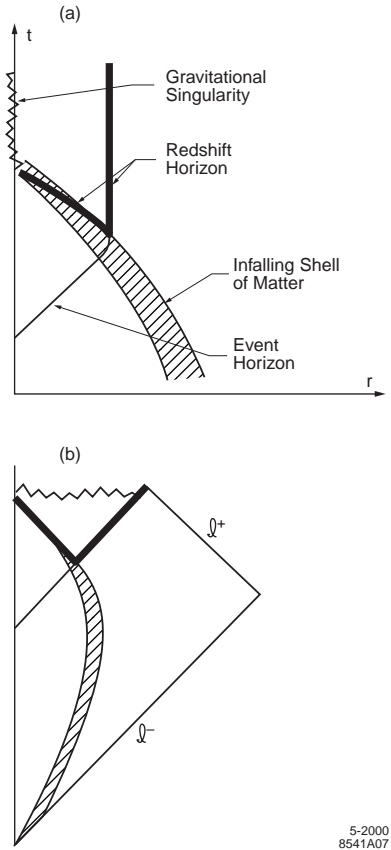}}
\end{center}
\caption[*]{(a)  Space-time history of a black hole created by a spherically
symmetric shell of infalling matter, in Schwarzschild coordinates.  Shown is
the Schwarzschild radius and the Penrose-Hawking event horizon, within
which no light ray can emerge to spatial infinity.  Also shown is the
``red-shift'', or ``apparent'' horizon which is pair-created by the infalling
matter and which separates the interior region with
$g_{00} < 0$ from the exterior region with $g_{00} > 0$.  (b)  The same picture,
in a Penrose diagram.}
\label{fig:7}
\end{figure}

If our existing theoretical formalism has finite bandwidth, then the
divergence in these quantities, which implies infinite bandwidth, may
signal a sensitivity to new-physics phenomena at the horizon. Without an
understanding of what the new physics is, there is no {\it necessity} that
something discontinuous happens to, say, the nature of the vacuum as the
horizon is crossed, but only that there is a reasonable {\it possibility} that
this might occur. I realize that this viewpoint cuts strongly against conventional
wisdom, because freely falling observers are not supposed to ``see" the
infinite bandwidth phenomena that the Schwarzschild surveyors see. And it
is also a necessity that the horizon which is considered here is not the
event horizon used by Penrose and Hawking in their global analyses, but
rather an apparent, or ``redshift" horizon. (See Fig.~\ref{fig:7} for a description
of the distinction I have in mind.)

\vspace{.5cm}

\eject

While I am not yet sure of my ground and recognize that most experts do
not share my doubts, I still find the Schwarzschild horizon a potential
frontier, something akin to the frontiers posed by the Planck and
cosmological distance scales. But maybe that is just my own shortcoming.

\section{The Fate of QCD}

As we have mentioned already, quantum chromodynamics (QCD), the theory of
the strong force, is arguably the most comprehensive of the quantum field
theories in use today, and in principle is a consistent closed theory with
infinite bandwidth. In practice, there are two major branches of QCD,
short-distance and long-distance. The former is dominated by a
Feynman-diagram approach. It is basically perturbative in nature, although
in practice large sets of diagrams need to be summed in order to attain
the needed accuracy. And the short-distance limit has a host of
applications and is of immediate relevance for all new-physics searches at
the high energy frontier. For all these reasons there has been and will
continue to be a large investment of effort in this area~\cite{ref:r}.

The large-distance limit, ``soft QCD", has to do with hadron structure and
vacuum structure. The distance scale ranges from above 10$^{-13}$ cm to a
little below 10$^{-14}$ cm, essentially the size scale associated with
ordinary hadrons. Here perturbation theory cannot be reliably used, and
therefore the theory is much more challenging. Many open questions remain,
the most prominent being a full description of the phenomenon of quark
confinement.

The vacuum structure of QCD is especially rich. At moderate distances
there is the challenge of understanding the role of instantons. At large
distances, there exists a ``chiral condensate", emergent from the
spontaneous breaking of the approximate strong-interaction chiral
symmetry. And the QCD phase diagram needs explication; at present there is
a lot of progress in the theory~\cite{ref:s}. And the exploration of heavy ion
collisions at CERN and soon at the RHIC ion-ion collider at Brookhaven~\cite{ref:t}
makes the experimental situation also a dynamic one.

At short distances hadrons are described in terms of the pointlike
quark-gluon, ``parton" degrees of freedom. Much is know about the momentum
spectrum of these partons, viewed in reference frames where the parent
hadron has very high momentum. But precious little is understood about the correlations
between the partons. For example, how they are distributed in the
transverse plane is still a serious issue, and even the multiplicity
distribution of the partons is not established. And the nature of the
low-momentum tail of the distribution, the ``wee" partons of Feynman,
remains an active and unresolved issue.

Ordinary collisions at the very high energies of the LHC are another
frontier. So many ``wee" partons participate simultaneously in the typical
central proton-proton collision that the complexity of the dynamics, {\it e.g.}
the number of relevant constituent collisions per proton-proton collision,
exceeds what will be dealt with in gold-gold collisions at RHIC.
Diffractive processes comprise another difficult and unresolved subfield,
one  which
increases in prominence with increasing energy.  It is a
``shadowy" topic which probably involves large-distance QCD concepts in an
essential way\cite{ref:abc}.

Even the venerable subject of hadron spectroscopy ought to have a rich
future. A modern ``electronic bubble chamber", built to exceed the classic
bubble chamber's acceptance and resolution, could improve the statistics
and data quality of all the old resonance-physics topics by a millionfold.
Many anticipated resonant states of hadrons, especially those made
primarily of gluons, remain to be discovered and carefully studied.

I prefer to label this whole subfield ``Hadron Physics", in analogy to
atomic, molecular, and nuclear physics. The goals of those subfields are
quite analogous to the goals of hadron physics---namely to study the
internal structure of their building blocks, as well as to study their
``vacua", namely the extensive, condensed-matter structures built from
those constituents.

These subfields, especially atomic and nuclear physics, used to be within
the mainstream of elementary particle physics. As the energy scale of
interest to particle physics increased, each evolved into a distinct
discipline. And I think this is already happening to hadron physics.
Relatively little attention is paid nowadays by mainstream particle
physicists to the subfield of hadron physics. And a great deal of the
subject matter is now appropriated by the nuclear physics community, even
though what they are doing can hardly be called nuclear physics.

While this evolution is basically a natural one, there is a new problem
which was not faced in the previous examples. Much of the experimental
program of hadron physics needs to be done via high energy collisions,
within the facilities built and exploited by particle physicists. It
is not easy for hadron-physics initiatives to be recognized and to receive
the necessary priority rating, whether from laboratory managements,
program committees, funding agencies, or the existing peer review system,
when a program dedicated to hadron structure is put in direct competition
with major high-energy physics initiatives. I believe that in order for
such hadron-physics initiatives to be viable, there must be institutional
changes at all these levels, which recognize the complementary nature of
hadron-physics research, and which provide a certain degree of guaranteed
access to the high-energy facilities in return for an appropriate
contribution to the operating costs of the facilities~\cite{ref:u}.

\section{Remarks}

In summary, elementary particle physics in the next century should
continue to be full of progress and full of vitality. While there is a
slowing of the pace, it is still the case that within the next decade we
should already witness a major turning point, namely a much better
understanding of the problem of mass and of the mechanism of electroweak
symmetry breaking. The way in which that problem is answered will have much to do
in shaping the nature of the experimental program beyond.

The vitality of theory nowadays is focused on the superstring ideas,
which continue to generate new ways of looking at the fundamental problem
of going beyond the standard model. The only problem with this activity is
its remoteness from data. This in itself causes me no problem.
But if it leads to an indifference of theorists toward the
data-driven side of the field, then I will have a problem.  

When new data
appears at a relatively slow rate, ideas and interpretations tend to
ossify, and it sometimes becomes harder not only to think in new ways, but
also to maintain a diversity of approach, which is always an essential
element of the scientific endeavor. The existence of the Standard Model
does not imply the existence of a standardized anticipation of the future.
The only thing that deserves institutionalization is doubt. This problem
of maintaining diversity of approach afflicts both experiment and theory,
and if I have any concern about how the field is developing, it is about
this point I worry the most.

In this respect, the second-tier initiatives such as the ``factories", and
new generations of non-accelerator facilities are important to encourage.
And I would add QCD and hadron-physics initiatives to this list as well. 
There is very much to do, most of which is relatively accessible provided
the resources are made available. And last but not least, the long range
problem of reaching extremely high energies should not be neglected,
implying good support for advanced accelerator R\&D.

\section{Acknowledgments}

It is a pleasure to thank the organizers of this conference for the
excellent program and fine hospitality extended to us all. I also
thank Pisin Chen, Leon Lederman, Chris Quigg, and Leonard Susskind
for critical comments and advice.

\end{document}